 \documentclass[preprint]{aastex}

\newcommand{\mR}{$m_{ROTSE}$ }
\newcommand{\iv}{I$_{var}$ }

\slugcomment{Accepted for publication in April 2000 AJ}

\shorttitle{ROTSE All Sky Surveys for Variable Stars I: Test Fields}
\shortauthors{Akerlof et al.}

\begin{document}


\title{ROTSE All Sky Surveys for Variable Stars I: Test Fields}


\author{C. Akerlof\altaffilmark{1}, S. Amrose\altaffilmark{1}, 
	R. Balsano\altaffilmark{2}, J. Bloch\altaffilmark{2}, 
	D. Casperson\altaffilmark{2}, S. Fletcher\altaffilmark{2}, 
	G. Gisler\altaffilmark{2}, J. Hills\altaffilmark{2}, 
	R. Kehoe\altaffilmark{1}, B. Lee\altaffilmark{3}
	S. Marshall\altaffilmark{4}, T. McKay\altaffilmark{1}, 
	A. Pawl\altaffilmark{1}, J. Schaefer\altaffilmark{1}, 
	J. Szymanski\altaffilmark{2}, J. Wren\altaffilmark{2}}


\altaffiltext{1}{Department of Physics, University of Michigan,
    Ann Arbor, MI, 48109}
\altaffiltext{2}{Los Alamos National Laboratory, Los Alamos, NM, 87545}
\altaffiltext{4}{Lawrence Livermore National Laboratory, Livermore, CA, 94550}
\altaffiltext{3}{Fermi National Accelerator Laboratory, Batavia, IL, 60510}


\begin{abstract}
The ROTSE-I experiment has generated CCD photometry for the entire
Northern sky in two epochs nightly since March 1998. These sky
patrol data are a 
powerful resource for studies of astrophysical transients. As a 
demonstration project,
we present first results of a search for periodic variable stars derived 
from ROTSE-I observations. Variable identification, period determination, and 
type classification are conducted via automatic algorithms.
In a set of nine ROTSE-I sky patrol fields covering
$\sim$2000 square degrees we identify 1781 periodic variable stars with 
mean magnitudes
between $m_v$=10.0 and $m_v$=15.5. About 90\% of these objects are newly 
identified as variable. 
Examples of many familiar types are presented. All classifications for this
study have been manually confirmed. 
The selection criteria for this analysis have been conservatively defined,
and are known to be biased against some variable classes.
This preliminary study includes only
5.6\% of the total ROTSE-I sky coverage, suggesting that the full ROTSE-I
variable catalog will include more than 32,000 periodic variable stars.

\end{abstract}




\section{Introduction}

The Robotic Optical Transient Search Experiment\footnote{For more 
details see \url{http://www.umich.edu/$\sim$rotse}}
 (ROTSE) is a collection of 
instruments designed to search for astrophysical transients, especially
those associated with Gamma-ray Bursts (GRBs). Along with observations 
triggered
by external sources, ROTSE instruments perform regular patrol observations.
In particular, the ROTSE-I instrument has been imaging the entire sky 
visible from New Mexico in two epochs nightly since March 1998. These 
observations provide a unique opportunity to perform uniform all-sky surveys 
for variable stars.

All-sky observations of variable stars, even at comparatively bright 
magnitudes, have important advantages over narrow field searches. 
Variables found in 
uniform all-sky surveys are ideal for performing studies of galactic 
structure. They can detect structure in the galaxy on all scales, while 
avoiding the confusion that substructure creates in pencil beam surveys.
All-sky surveys are more sensitive to
intrinsically faint classes of disk variables, such as Delta
Scuti stars and main sequence contact binary systems. 
They are also very useful for finding
rare, intrinsically bright variables, such as red 
supergiant variables \citep{fea80}. 
Finally, all sky surveys at relatively bright magnitudes identify complete
samples of nearby variables, ideal for detailed study by other techniques
such as parallax measurement, x-ray observations,
and high resolution spectroscopy. This is especially important as we prepare
for the era in which Full-sky Astrometric Mapping 
Explorer\footnote{see \url{http://aa.usno.navy.mil/FAME}}
and the Space Interferometry 
Mission\footnote{see \url{http://sim.jpl.nasa.gov/index.html}} 
will be able to use these objects to
directly calibrate the distance ladder.

We present here first results from analysis of ROTSE-I sky patrol data. For
this analysis we have concentrated on the study of periodic variable stars.
A discussion of aperiodic transients will be presented in a subsequent 
publication. In 
the following sections we describe relevant details of the ROTSE-I system,
data reduction, variable identification, phasing, and automatic 
classification. This is followed by a description of some general properties
of the variables discovered.
We conclude with a discussion of what we expect from future ROTSE variable 
catalogs.

\section{Observations}

The ROTSE-I instrument consists of four Canon 200mm f/1.8 lenses, each
equipped with a thermoelectrically cooled CCD camera 
incorporating a Thompson TH7899M CCD.
The $14~\mbox{$\mu$m}$ pixels of these CCDs subtend 14.4'' at this 
focal length. Each of
these four assemblies has a field of view 8.2\degr x 8.2\degr. All four
optical assemblies are co-mounted on a single rapidly slewing mount. Pointing
offsets between the four optical assemblies allow the instrument 
to cover a combined
16$^\circ$x16$^\circ$ field of view. The Canon lenses provide a 
point spread function which has 
a typical full width at half maximum
of about 20''. As a result stellar images are only moderately 
undersampled. To maximize sensitivity to GRB optical transients, ROTSE-I
CCDs are currently operated without filters.

All ROTSE instruments are designed for completely automatic operation. 
The ROTSE-I
telescope is housed on the roof of a military surplus electronics enclosure.
It is protected by day and in bad weather by a clamshell cover which flips
completely out of the way during operation. Inside the ROTSE-I enclosure
are a set of five Linux PCs which operate the system. Four of these are 
dedicated to operation of the CCD cameras. The fifth is a master computer
which completely controls observatory operations. ROTSE instruments are 
currently installed at Los Alamos National Lab, near the LANSCE end station.
Deployment of these instruments to a 
dark site at Fenton Hill, west of Los Alamos, is contemplated for the 
near future.

At the beginning of the night the ROTSE master computer checks the 
weather status, which is measured by local monitoring
hardware. If conditions allow, it opens the clamshell and
waits for astronomical twilight. ROTSE has two principle observing modes;
patrol mode and trigger response mode. Most of the time is spent in
patrol mode. The large ROTSE-I field of view 
allows the entire celestial sphere to be tiled with a set of 206 field 
centers (see Figure \ref{patrol_map}). In its current New Mexico 
location, ROTSE-I can image 160 of these fields at various times of 
the year.
While in patrol mode, we successively image all available sky patrol fields. 
Occasionally (about once every ten days) an accessible GRB
trigger arrives through the GCN system \citep{bar98}. 
When this occurs, ROTSE immediately
interrupts patrol observing and obtains burst response data. Details of
ROTSE studies of GRBs, including the first observation of optical
emission contemporaneous with a GRB,
can be found in \citet{ake99} and \citet{keh99}.

Sky patrol observations are conducted in the following manner. A list of 
available (elevation $>$20$^\circ$) fields is generated at the start of
a patrol.
During the night, the telescope slews to each patrol location and obtains
a pair of 80 s exposures. Exposures are taken in pairs to eliminate 
confusion caused by cosmic rays, satellite trails, etc. and to
allow robust detection of aperiodic transients. Paired observations
also prove extremely useful for detection of periodic variables, as described
below. 
On a typical night two patrol sequences are obtained, covering about 
18,000 square degrees of sky, and recording the brightness of 
$\sim 9 \times 10^{6}$ stars with four images taken in two epochs.
The time of each
observation is recorded with an accuracy of 20 ms. Maintainance of an accurate
time standard through the system is accomplished by use of the Network 
Time Protocol, and is required to support ROTSE's GRB mission.

Instrumental calibrations required for reduction of ROTSE 
data are also obtained
automatically. A sequence of 12 dark frames is recorded
during the night. These dark frames are median 
averaged to provide a global dark for each night. Since the cameras are
TE cooled, this correction is primarily
important for removing the small number
($<<1\%$) of pixels which have high dark current rates. Since ROTSE-I 
pixels are
so large, they have relatively high sky rates. As a result, patrol 
observations are sky noise limited. We obtain flatfields from
night sky images 
by median averaging
a single observation of each patrol field. As there are typically 
90 fields observed in at least one epoch, 
this process yields very good, stable flats. Flatfield corrections
for ROTSE-I are dominated by vignetting in the lens, which amounts to about
a 40\% loss of sensitivity at the CCD corners.

Regular patrol observations by ROTSE-I began in March 1998. These early
observations utilized 25 s patrol exposures, reducing the sensitivity, but
increasing the number of observation epochs. In March 1999 exposure lengths
were increased to 80 s. As of November 1999
more than 2.6 terabytes of imaging data have
been obtained, a total of about 430,000 images. For this analysis we have 
selected 9 of the 160 patrol fields
for which ROTSE-I data exist. We have analyzed all observations of these fields
obtained from March 15, 1999 to June 15, 1999. The number of available epochs
varies by field from $\sim$40 to $\sim$110. These data includes 
about 40\% of the available time coverage for these fields, which constitute
5.6\% of the ROTSE-I sky coverage.

\section{Data Analysis and Calibration}

Except for automatic generation of darks and flats, ROTSE-I sky patrol 
analysis for this study was conducted offline. Online analysis of some
data began in August 1999, and we expect to begin online analysis of
all data in Spring 2000.
Data analysis begins with frame correction. Median dark images are subtracted
from each sky exposure. Images are then corrected for variable system response
by application of the night sky flats described above. These corrected images
are then reduced to object lists by the SExtractor \citep{ber96} 
package. Since ROTSE-I data is heavily dominated by stars we retain only
a small set of the available SExtractor outputs: position, size,
magnitude, and
magnitude errors. The remainder of the analysis, from calibration through
automatic variable classification, is carried out with a series of IDL 
routines generated at the University of Michigan over the last several years.

Calibration for the ROTSE-I data is accomplished in a somewhat unusual manner.
Each camera in the ROTSE-I array has a 64 square degree field of view.
This implies that within each image there are typically 1500 Tycho 
\citep{hog98} stars. The Tycho catalog is derived from Hipparcos
observations, and includes both highly accurate astrometry and two-color
(B and V) space-based CCD photometry. Most Tycho stars are fainter 
than the $m_{V}\simeq$ 9.0 
saturation limit of ROTSE-I observations. Astrometric and photometric 
calibrations are based on these stars. The availability of large
numbers of well measured calibration stars in each image is a remarkable
resource, both for calibration and monitoring of data quality.

\begin{figure}
\plotone{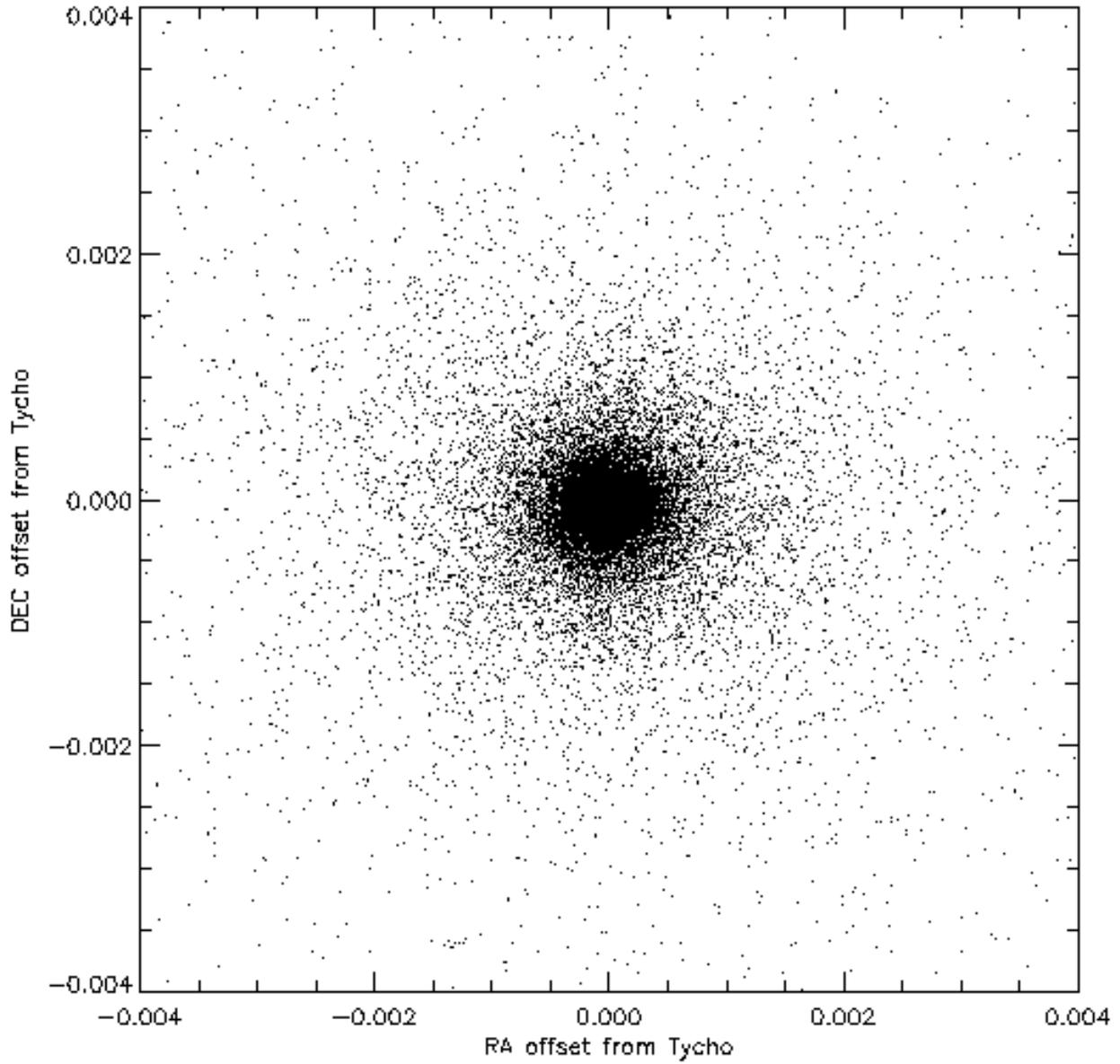}
\figcaption[astrometric_accuracy.ps]
{This figure shows astrometric residuals between Tycho and 
ROTSE positions for all 28740 jointly observed objects. The full
size of the plot is $\pm$1 ROTSE-I pixel. RMS residuals for these
bright ($m_{V} <$ 11.5) stars are about 0.1 pixel.
\label{astrometric_accuracy}}
\end{figure}

Astrometric calibrations are accomplished by identifying ROTSE-I objects
with Tycho stars through a triangle matching routine. The transformation
of CCD (x,y) to (ra,dec) is accomplished through a third order polynomial 
warp. The resulting quality of the astrometric calibration can be tested 
by examining the residuals to this fit. Astrometric residuals for a typical
field are shown in Figure \ref{astrometric_accuracy}. 
Errors for bright ($m_{V} < 12.5$)
stars are $\sim$1.5'', about one tenth of a pixel.


Photometric calibration is somewhat more complex. The Tycho catalog includes
only B and V photometry, and ROTSE-I images are obtained with unfiltered
CCDs. We use the Tycho B and V magnitudes to produce an empirically
predicted m$_{ROTSE}$ for each Tycho star:
\begin{equation}
m_{ROTSE} = m_{V} - \frac{m_{B}-m_{V}}{1.875}
\end{equation}
These color-corrected magnitudes are then used compared to ROTSE-I 
instrumental magnitudes
to set ROTSE-I zeropoints. The net
effect of this procedure is to place ROTSE observations onto a 
V-equivalent scale, in the sense that the average Tycho star has  
m$_{ROTSE}$=$m_V$. 

\begin{figure}
\plotone{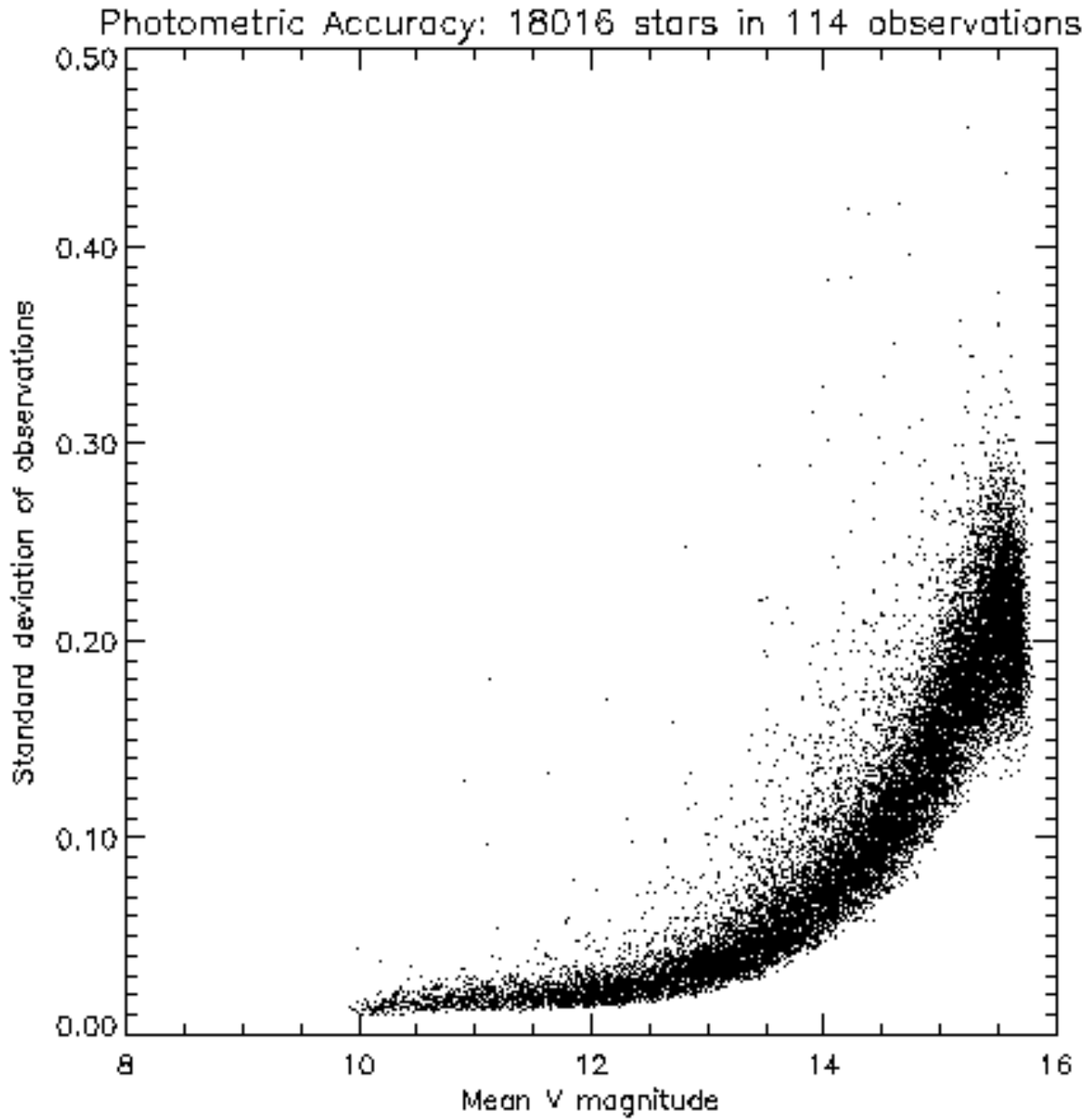}
\figcaption[photometric_accuracy.ps]
{This figure shows photometric residuals for 114 observations of 
about 18,000 objects over a period of four months. Errors range from about 
2\% (mostly systematic) at 10th magnitude to about 20\% (mostly 
statistical) at the magnitude limit of 15.5. Outliers are real variables. 
\label{photometric_accuracy}}
\end{figure}

While m$_{ROTSE}$ is clearly not a standard V magnitude, this 
technique provides very good, stable, zeropoints.
As a measure of the quality and stability of these
photometric
calibrations we present in Figure \ref{photometric_accuracy} the standard 
deviation
of 18016 stars observed in 114 different epochs over a period of four months.
The rms errors rise from $\sim$2\% at 10th magnitude to about 20\% at the
5$\sigma$ threshold around \mR=15.5. These errors are typical of the
fields included in this study. 

The presence of many Tycho stars in each image provides an excellent 
opportunity for monitoring data quality. The calibration routines 
generate summary outputs including both astrometric and photometric
fit residuals. This summary information is carefully examined before 
observations from a particular day are accepted and incorporated into the
overall light curve catalog.

Once calibrated object lists from each observation of each field are generated,
they must be collated into tables of light curves. This is done
using the calibrated object positions. A requirement is imposed that each 
object must appear in both observations of a ROTSE pair to be included
in the light curve measurement. This provides a strong veto against
cosmic rays, satellite glints, etc. The output of this process is 
an array of light curves for each field. These light curve tables then 
form the input to the subsequent process of variable identification and 
classification.

\section{Variable Identification}

To automatically detect variables we implement the technique of 
\citet{wel93} (the WS technique) as modified by 
\citet{ste96}. This method increases
our sensitivity to periodic variables by taking advantage of
paired observations. Both observations in a ROTSE-I pair are
taken within 2.5 minutes. Since this time is much shorter than
the period of the variables we seek, we expect the two observations
to record essentially the same magnitude. For variable objects, the residuals
from the comparison of each magnitude to the mean magnitude will
be correlated. For stable objects there is no correlation of residuals.
Products of these residuals will, for variables, generally 
be positive, while products of uncorrelated residuals may be either
positive or negative. As a result, the sum of these
products for paired observations will increase monotonically for 
periodic variables and cluster around zero for stable objects.

For each object with a light curve we calculate a variability index through a 
series of steps. The variations from the most obvious implementation
of the WS technique are all designed to make the measurement more robust
against the presence of a few bad observations. We begin by forming the
appropriately weighted products of residuals for each pair of observations.
\begin{eqnarray}
\delta_{1i} = \frac{V_{1i} - \overline{V}}{(\sigma_{V})_{1i}} 
	\nonumber \\
P_i = \delta_{1i}\delta_{2i} 
\end{eqnarray}
\noindent
where $\delta_{1i}$ is the uncertainty weighted residual 
between the first of a pair of observations of an object and
the mean magnitude of that object through all observations. $P_i$ is the 
product
of the residuals from a pair of adjacent observations. The mean magnitude
used in this calculation is based on the robust mean of \citet{ste87}. It is 
an iteratively weighted mean, where the weights are 
based on the residuals between each observation and the previously determined
mean. The uncertainty $\sigma_{V}$ used for this calculation is the 
photometric error calculated by SExtractor added in quadrature to
an assumed 2\% systematic error. We generate sums of these products of 
residuals to calculate the variability index
\begin{equation}
J_{ROTSE} = \sqrt{\frac{1}{n(n-1)}}
	\sum_{i=1}^{n}sgn(P_i)\sqrt{\mid P_i \mid} 
\end{equation}
\noindent
where n is the number of available epochs. This varies by field from
22 to 114, primarily because different time periods were
analysed for each field.
This form is simpler than that discussed by \citet{ste96} because all
ROTSE observations appear in pairs, and all pairs are equally weighted. 
As a last addition, we modify this index by the kurtosis measure of
\citet{ste96}. This parameter is designed to account for the fact that
many real variables have sinusoidal light curves. Magnitude measures 
for a sinusoidal variable cluster around the maximum and minimum values.
Those for a stable star with Gaussian errors cluster around the mean
value. As a measure of this we calculate
\begin{equation}
K_{ROTSE} = \frac{\frac{1}{N}\sum_{i=1}^{N}{\mid \delta_i \mid}}
	{\sqrt{\frac{1}{N}\sum_{i=1}^{N}{\delta_{i}^{2}}}} \nonumber
\end{equation}
\noindent
In this equation N represents the total number of observations (twice the
number of epochs) and $\delta_{i}$ is the residual from the mean of 
each observation.
This parameter has values K=1.0 for a square wave, K=0.90 for a sinusoid,
and K=0.798 for a stable object with Gaussian errors. When the residuals
$\delta_i$ are dominated by a single bad measurement with residual $\Delta$, 
this parameter becomes
\begin{eqnarray}
K_{ROTSE} \simeq \frac{\frac{\mid \Delta \mid}{N}}
	{\sqrt{\frac{\Delta^2}{N}}} \nonumber \\
	\simeq \frac{1}{\sqrt{N}}  
\end{eqnarray}
\noindent
Since this goes to zero in the limit of large N, the kurtosis index helps
to reduce the sensitivity to single bad observations. It also reduces our
sensitivity to variables with a low duty cycle, such as flare stars and
detached eclipsing binaries.
We combine this with the variability index $J_{ROTSE}$ to determine the 
final selection index
\begin{equation}
I_{var} = J_{ROTSE}K_{ROTSE}/0.798
\end{equation}
\noindent
This is equal to J when the residuals are Gaussian, slightly amplified for
sinusoidal light curves, and significantly reduced when a single bad 
measurement dominates the variability index J.

\begin{figure}
\plotone{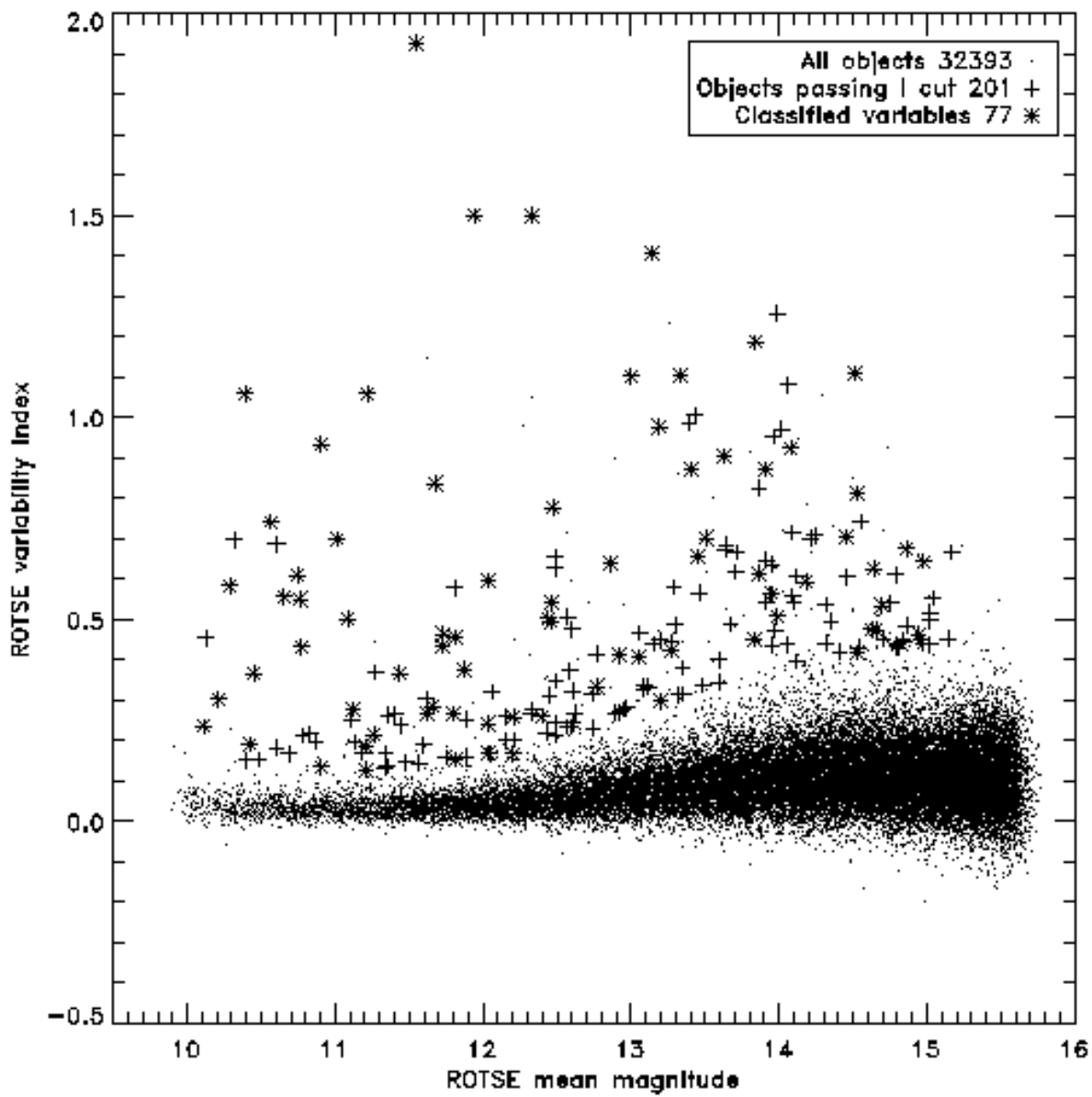}
\figcaption[iv_plot.ps]
{The ROTSE variability index $I_{var}$ 
calculated as described in the text is shown
as a function of mean magnitude. Objects which pass the cut are shown
as plus signs. Those finally selected as variable are stars. The points
with large values of the variability index which are not selected have
been rejected by the flare and deblending cuts described in the text. 
\label{iv_plot}}
\end{figure}

Variable candidates are selected by examining \iv for all objects which are
present in at least 11 epochs. This requirement is made because phasing
is ambiguous for objects observed in fewer epochs.
Figure \ref{iv_plot} shows the distribution
of \iv vs. \mR for a typical field. For this analysis variables are
selected to be those which have an \iv value more than 4.75 $\sigma$
above the mean value. In a typical camera (64 square degrees), this cut 
identifies between 40 and 100 candidate variables.

Two remaining cuts are made to clean the sample. Despite the 
robustness of \iv against objects whose light curves have just a few 
deviant pairs, there remain a small number of `flaring' objects per 
field which pass our \iv cut. Some of these objects are 
real variables such as detached eclipsing binaries and flare 
stars. Since we have little information on their variability, we opt
to remove them from this analysis and reserve them for further study.
Such `flare' objects are defined as objects for which there are at
most two pairs of observations falling in the magnitude range from
0.5 to 1.0 times the absolute value of the maximum residual. This cut
removes about 25\% of candidate variables.

Finally, we make a cut designed to remove problems with deblending. The
SExtractor deblending algorithm works well for these stellar images, but
there are some cases where stars are sufficiently close to one another
that they are sometimes deblended, and sometimes blended. This creates
a very characteristic light curve in which the measured magnitude 
shifts between two distinct values. To detect these automatically, we
first generate a list of ROTSE observed stars (if there are any) within a 
few pixels
of each variable. We then extract a subset of the observations in which
the candidate variable is found and its close neighbor is not. These
are candidate `blend' measurements. We compare the mean and standard 
deviation of the `blend' and `non-blend' measurements. If the means
are separated by more than two standard deviations, the object is considered
a deblending problem and removed from the variable list. The effect of
this cut is strongly dependent on crowding (and hence on galactic latitude).
It removes between 5\% and 30\% of candidate variable objects.

The output of this process is a set of objects which show moderate evidence
for variability. At this point we allow the 
inclusion of some false variables, as we wish to maintain sensitivity
to real variables of the lowest possible amplitudes. We use robust
evidence of periodicity as our final restriction. This list of 
variable candidates includes 7396 objects, drawn from a total of 
917,266 which pass the 11 epoch cut.

\section{Variable Confirmation and Classification}

Variable classification proceeds in two steps. First, an attempt is made
to fit each light curve to a third order polynomial over the full period 
of observation. Since these observations cover about 100 days, this fit
can be quite good for a variety of long period variables, and is 
used to flag long period objects. For these objects
the polynomial fit parameters are used to derive an amplitude
(maximum variation within the observation window) and a measure
of goodness of fit. Of the 7396 variable candidates, 739 are identified
as long period variables by this technique.

The second step is to attempt phasing of all objects which pass the \iv
cut. For this purpose we
use a cubic spline method described in detail in \citet{ake94}. This 
technique provides best fit periods, period error estimates, and spline
fit approximations for object light curves. The general quality of 
these spline fits is illustrated in the sample light curves described
below. In total 1195 of our 7396 variable candidates are successfully phased. 
The remainder (5462) are mostly near the 
detection threshold. Additional ROTSE data will be required to determine
how many of the remaining candidates are real variables.

With phased light curves in hand, the classification process begins.
Automatic classification is based on period and light curve shape, as 
quantified by the spline fit. We begin by Fourier analyzing the 
spline fit to the light curve:
\begin{equation}
\Delta_{m}(t) = \sum_{i=0}^{255}p_{i}cos(\frac{2\pi t}{\Gamma})
\end{equation}
\noindent
Where $\Delta_{m}(t)$ is the phased light curve. Classification then 
relies on the derived period $\Gamma$, ratios of the Fourier coefficients:
\begin{center} $
\begin{array}{c}
r_{1}=\frac{p_{2}}{p_{1}} \\
r_{2}=\frac{p_{3}}{p_{1}} \\
r_{3}=\frac{p_{2}}{p_{3}} 
\end{array}$
\end{center}
and the sign of the largest deviation from the mean. Cuts based on these
Fourier power ratios are referred to below as `ratio cuts'. 
In addition, most 
classifications require that the largest power in the Fourier series is
in the first term. This implies that the phased light curve has one 
cycle per period. The cuts used here were selected through examination of 
the parameters of a subset of manually classified objects. We expect them 
to evolve in future ROTSE variable studies.

Classification for the purposes of this study is 
confined to 8 classes: RRab, RRc, Delta Scuti, Cepheid, Contact Binaries,
Eclipsing, Mira, and long period variable. 
The classifications given here should be 
considered preliminary. We are aware that there is not necessarily 
complete correspondence
between our classifications and more traditional definitions of these classes.
We refer to our classes as RRAB, RRC, DS, C, E, EW, M, and LPV in 
what follows.

RRAB stars were the original target of this project. Although most RRABs have
periods from 0.3-0.9 days, some have been detected with periods as long as
2 days. Therefore the period range is defined as 0.3-2.0 d. 
Classification within this range is based on ratio cuts. RRAB stars are 
characterized by asymmetric light curves with a rapid rise followed by a
slower decay. As a result we search for light curves which are not 
sinusoidal, but have substantial contributions from higher harmonics.
RRAB stars are selected to have:
\begin{center} $
\begin{array}{ccc}
0.08 < r_{1} < 1.0, & 0.01 < r_{2} < 1.0, & r_{3} > 0.6
\end{array} $
\end{center}
The RRC, DS, and C types are quite similar in the sense that all have
more or less sinusoidal light curves. 
The primary distinction between these classes is period range. We begin by 
requiring that the contribution from higher harmonics be small.
All three classes share the same ratio cuts:
\begin{center} $
\begin{array}{cc}
r_{1} < 0.16, & r_{2} < 0.024
\end{array} $
\end{center}
Those phased objects which pass these cuts we classify as DS ($\Gamma< 0.2$d),
RRC ($0.2$d$ < \Gamma < 1.0$d), or C ($1.0$d$ < \Gamma < 50.0$d). 

\begin{figure}
\plotone{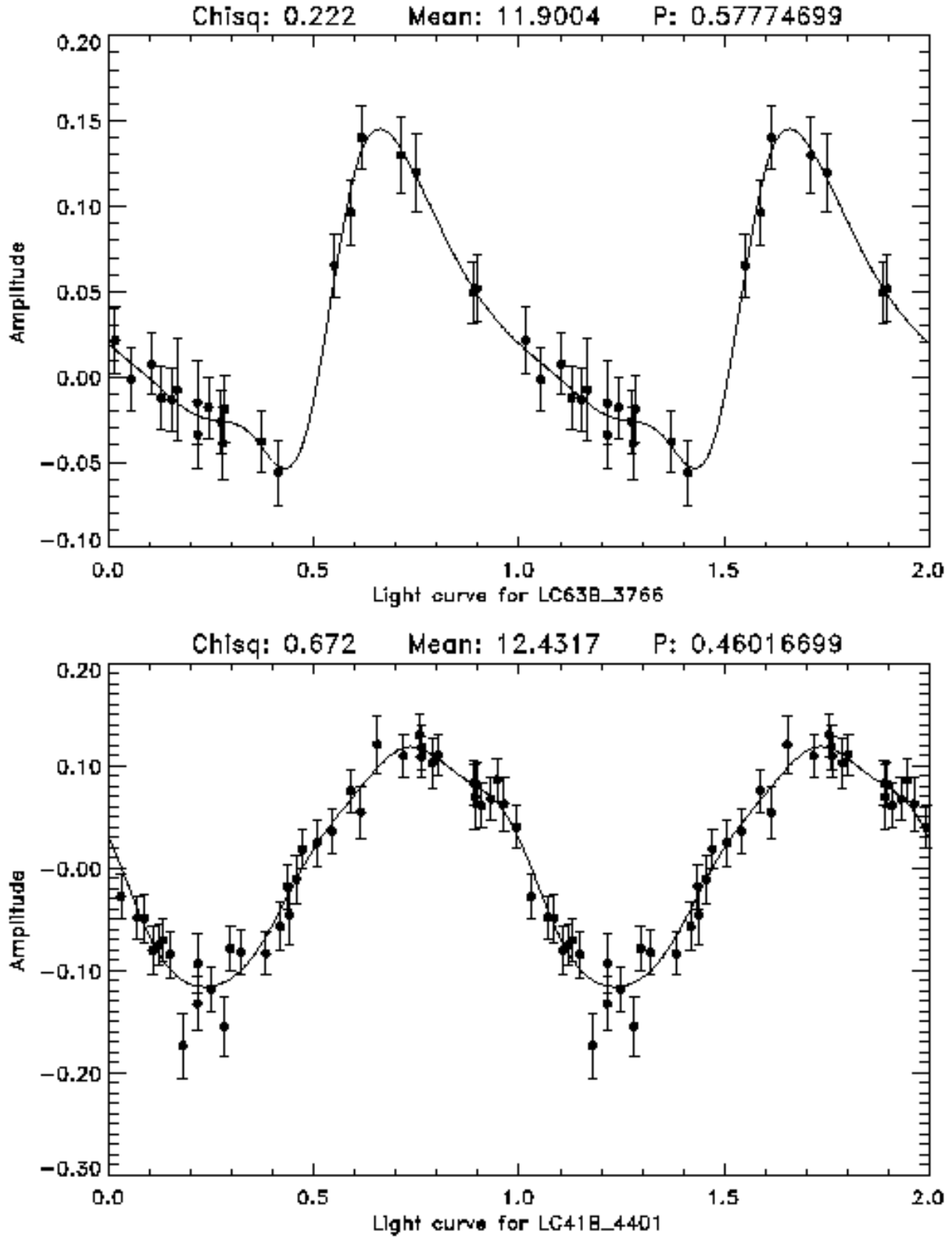}
\figcaption[rrab_rrc_comparison.ps]
{Light curves for an example RRAB (top) and RRC (bottom) in the overlap
region of period and amplitude space. The RRAB exhibits a steep rise
compared to the nearly sinusoidal RRC light curve shape.
\label{rrab_rrc_comparison}}
\end{figure}

This RRC classification
overlaps with the low end of the RRAB period and ratio space. This is not 
an enormous problem, as the majority of RRC stars have periods less
than 0.4 d. To accommodate the region of overlap we tighten the RRC ratio cuts
to $r_{1} < 0.08$ and $r_{2} < 0.01$ for the range $\Gamma > 0.4$ when the 
amplitude $>$ 0.35 m. This accounts for the only apparent difference 
between RRAB and RRC objects in this range; the RRC stars have lower 
amplitude and a light curve too sinusoidal to be called RRAB. A comparison of 
RRAB and RRC light curves in this period and amplitude range is 
given in Figure \ref{rrab_rrc_comparison}.

Eclipsing objects include the EW (contact) and E (detached) types. There are
no period restrictions for selection of eclipsing systems. EW objects are
selected as those with 
\begin{center} $
\begin{array}{cc}
0.04 < r_{1} < 0.2, & 0.007 < r_{2} < 0.04
\end{array} $
\end{center}
Since these objects tend to overlap with RRC and DS objects, the 
distinguishing feature becomes the sign of the greatest deviation. For
EW stars, the greatest deviation is always less than the mean. For RRC and 
DS stars the greatest deviation tends to be brighter than the mean.
Systems which receive the E classification also require a negative greatest
deviation. E type systems are treated in two sets, those
with eclipses of equal depth, and those with eclipses of different depths.
This second case is the only class which we allow to have two cycles per 
period. For those objects which have minima of equal depth, the real orbital
period of the system is twice the photometric period. Periods for
equal depth E  and EW type systems are corrected for this effect once 
classification is complete.

As a final step for this classification, each light curve is checked by
visual inspection, 
allowing for small adjustments to classification. In addition, each 
light curve and its 
classification was manually graded for quality from 10 (excellent) to 1
(marginal). Experience gained from this hand-scanning convinces us that
we will be able to essentially automate classification by implementing
further techniques like those described by \citet{ruc97}.

\section{Results}

\begin{figure}
\plotone{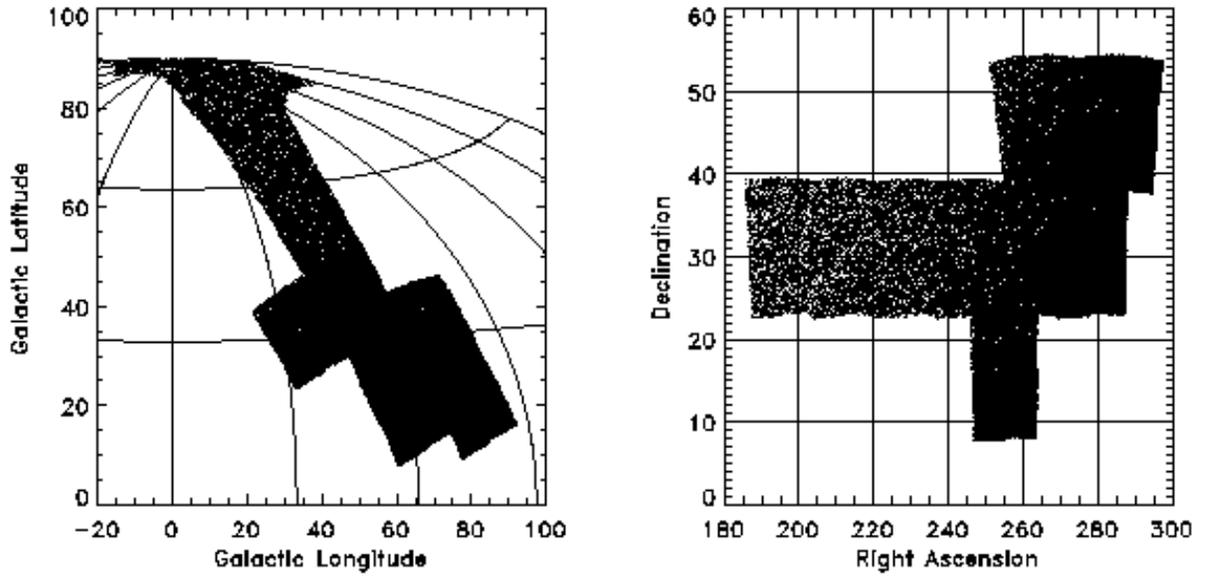}
\figcaption[rsv1_sky_coverage_plot.ps]
{The locations of the nine ROTSE-I fields analyzed for this project are shown
in galactic and celestial coordinates. The points represent all 77481 stars
brighter than 12th magnitude seen in these fields. 
\label{rsv1_sky_coverage_plot}}
\end{figure}

In this preliminary study we have analyzed $\sim$2000 square degrees of 
ROTSE-I 
sky patrol data in an effort to assess the usefulness of these data for
detection of periodic variables. The locations of these fields in galactic
and celestial coordinates are shown in Figure \ref{rsv1_sky_coverage_plot}.
A primary conclusion of this work
is confirmation that ROTSE-I data form an excellent resource for
the discovery and classification of periodic variables. We have discovered
a total of 1781 periodic variables, 89\% of which are not included in
the General Catalog of Variables Stars \citep{kho98} (based on position 
matching within a 28.8'' aperture). This reiterates the
long standing assertion \citep{pac97} that many variable stars brighter
than 15th magnitude remain to be discovered. We refer to this catalog
as the ROTSE Survey for Variables 1 (RSV1).

\begin{figure}
\plotone{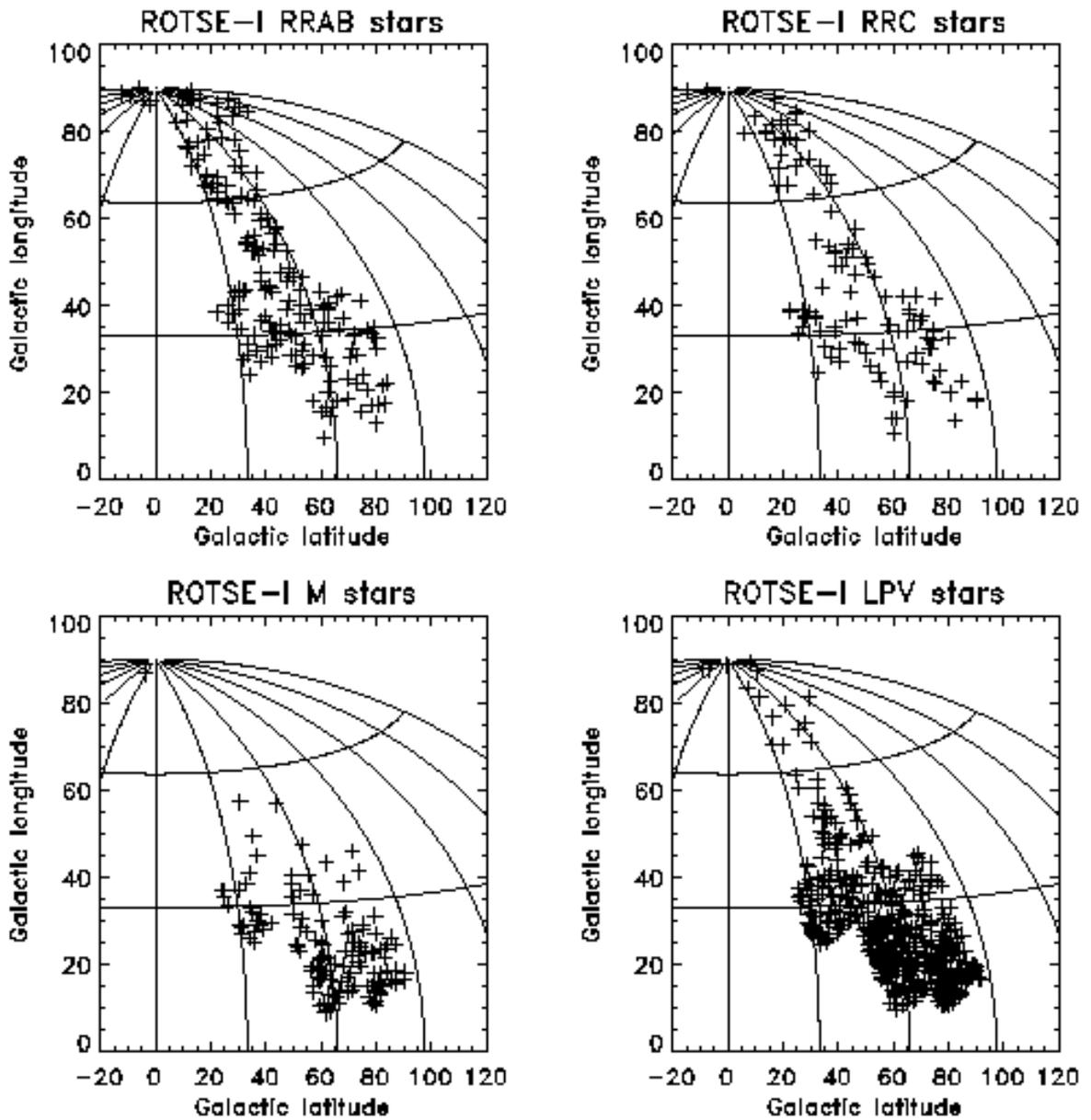}
\figcaption[sky_dist.ps]
{The distribution of ROTSE-I variables of four types in an Aitoff projection
of galactic coordinates. \label{sky_dist}}
\end{figure}

Some general properties of
the objects discovered are presented below. Example light curves 
are presented for each variable class to give an idea of data quality.
The distribution of several classes of objects on the sky is shown in 
Figure \ref{sky_dist}. The basic object list is presented in Table 
\ref{example_table}.
The entire catalog, along with light curves, is available online through
\url{http://www.umich.edu/$\sim$rotse}. 

Accurate determination of our variable detection efficiency 
is a complex exercise. Since detection efficiency is a strong function
of period, amplitude, and light curve shape, it must be determined 
seperately for each variable class. This can be done correctly
only after a complete understanding of the period, amplitude, and light
curve shape distributions in the observing bands is obtained. These
studies will be reported for each variable class in future publications.

For the moment we make a simple estimate of detection efficiency by 
measuring the fraction of GCVS stars recovered here. For this purpose we 
have selected GCVS stars within our survey area with maxima less than 
$m_{V}$=10 and minima greater than $m_{V}$=13.
Within this sample we recover more than 80\% of the RR Lyrae and Delta
Scuti stars. Our efficiency for eclipsing types, as expected, varies
strongly with type, from $\sim$70\% for close binaries to $\sim$30\%
for detached systems. All 52 known Miras within the ROTSE-I magnitude range
are recovered. Our lowest recovery rates are for GCVS types
L and SR (the slow irregulars and semiregulars) at about 37\%. This low
efficiency is due to a combination of variability timescale, amplitude,
and aperiodicity.

\subsection{RR Lyraes}

RR Lyrae stars are extremely useful as distance indicators and tracers 
of the structure of the Milky Way. They are attractively easy to identify
and measure well. As a result they were the original target of this 
investigation. With a typical $M_V$=0.74 \citep{fer98}, RR Lyraes 
are detectable by ROTSE-I from 0.7 to about 7 kpc.
Among all the variable types which we classify, our
classification of objects as RR Lyrae stars is most secure.

\begin{figure}
\plotone{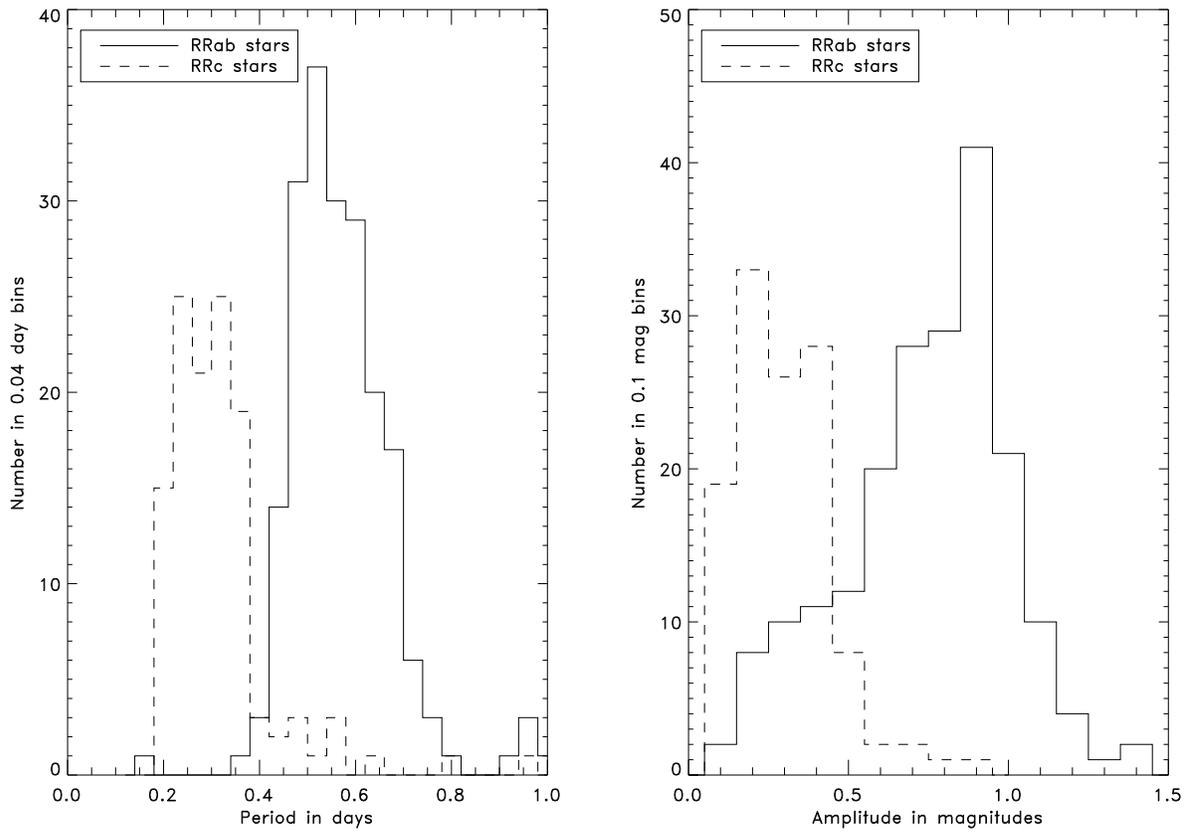}
\figcaption[rrl_period_mag_plot.ps]
{This figure shows the period and magnitude distributions for all the 
discovered RRAB and RRC type stars \label{rrab_per_mag}}
\end{figure}

\begin{figure}
\plotone{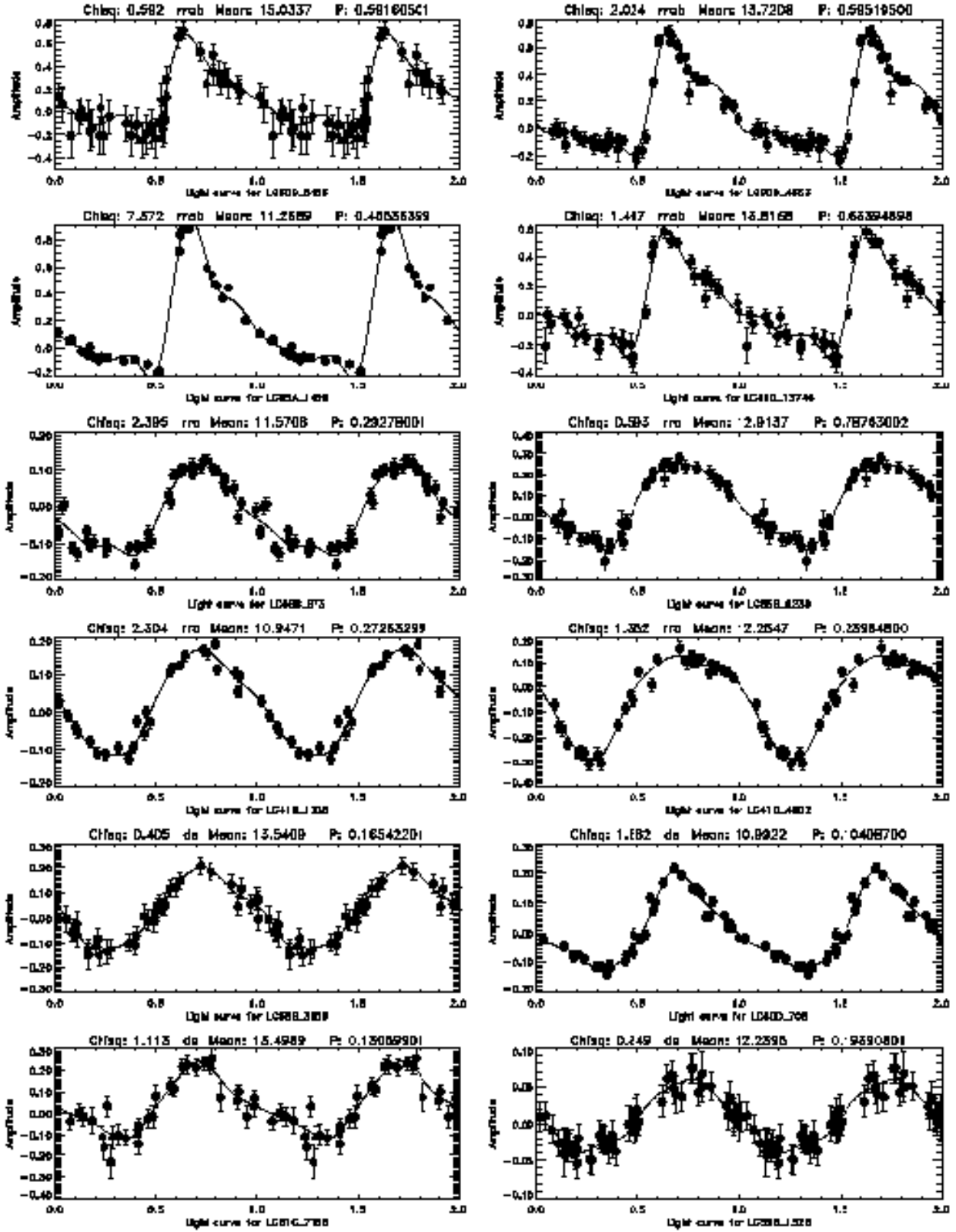}
\figcaption[rrab_rrc_ds_lcplot.small.ps]
{Example phased light curves for a random sample of ROTSE RRAB stars
(top four), ROTSE RRC stars (middle four), and ROTSE DS stars (bottom
four). Included above each plot are periods and 
mean magnitudes. \label{rrab_rrc_ds_lcplot}}
\end{figure}

In this preliminary survey, we identify 186 RRAB stars, 126
of which are newly discovered. As these stars have large amplitude and
distinctive variability, it is not surprising that a relatively
large fraction (32\%) are previously known. The period and amplitude
distributions for these stars are shown in  Figure \ref{rrab_per_mag}.  
All ROTSE RRABs which appear in the GCVS
are classified there as either RRAB (55 of 60) or 
RR (5 of 60). Sample light curves
for RRAB stars are shown in Figure \ref{rrab_rrc_ds_lcplot}.

\begin{figure}
\plotone{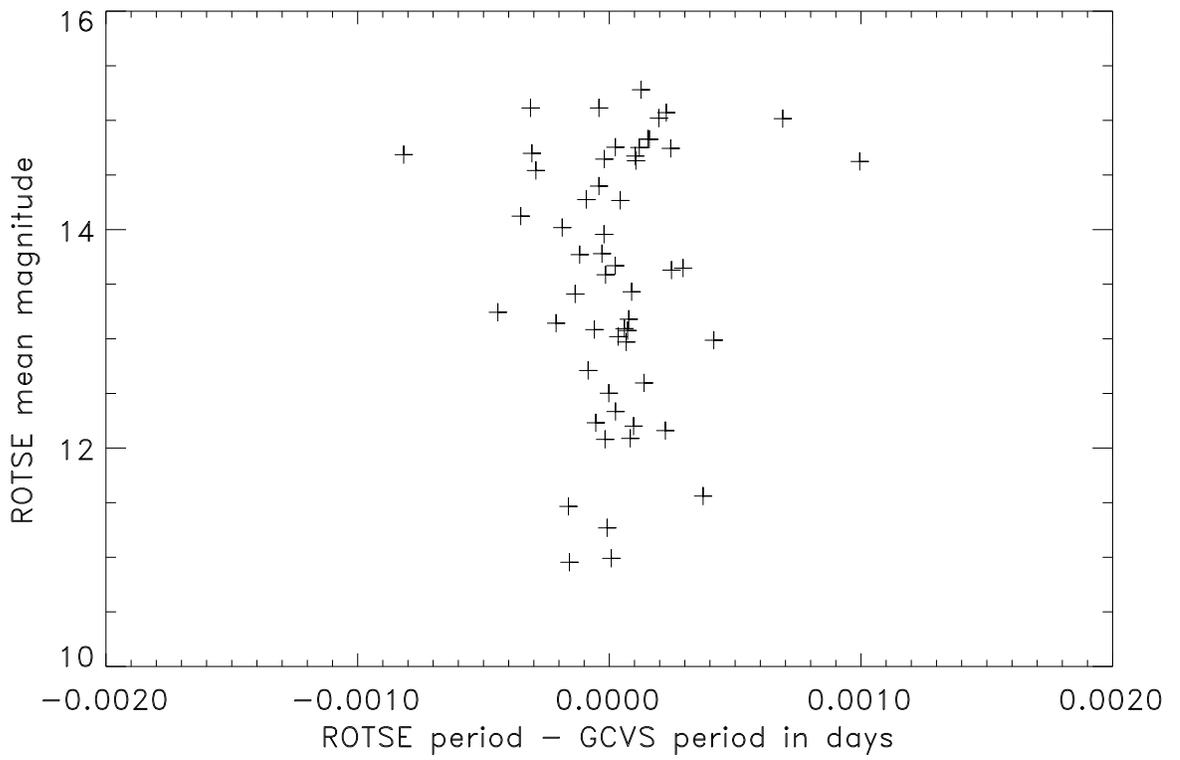}
\figcaption[rrl_period_compare.ps]
{The difference between GCVS and ROTSE derived periods is shown as
a function of mean magnitude for the overlap sample of 57 RRab stars.
\label{rrab_per_compare}}
\end{figure}

Since the RRAB sample has a large overlap
with the GCVS, we have made direct comparisons of ROTSE and GCVS derived
periods for these objects. The dependence of this period difference on
\mR is shown in Figure \ref{rrab_per_compare}. There are 57 overlapping
stars with measured GCVS periods. For two of these ROTSE measures
periods substantially different from the GCVS period. In one case the
ROTSE period provides an enormously better fit to the ROTSE data, suggesting
that either the GCVS period is in error or the period of this object 
has changed. In the second the ROTSE phase
coverage is incomplete. Examination of additional ROTSE data for this object
shows much better agreement with the GCVS period. For the remaining 
55 overlap stars the RMS period error between ROTSE and the GCVS is 
0.00026 d. Typical
period errors for ROTSE determinations are 0.00012 d, so these offsets are
probably dominated by GCVS period errors.

In addition, we identify a total of 113 RRC stars, 104 of which are newly 
identified. The large fraction of newly identified RRC stars is not 
surprising given their relatively small pulsation
amplitudes. The classification of
these stars as RRCs and not, for example, contact binaries is dependent
on hand scanning of the light curves. Of the nine objects known in the
GCVS, seven are classified there as RRC, two as EW. Visual examination of 
the light curves supports the 
classification as RRC in both cases of disagreement. The period and
magnitude histograms for these objects are also given in Figure 
\ref{rrab_per_mag}. Examples of light curves for these objects are 
presented in Figure \ref{rrab_rrc_ds_lcplot}.

\begin{figure}
\plotone{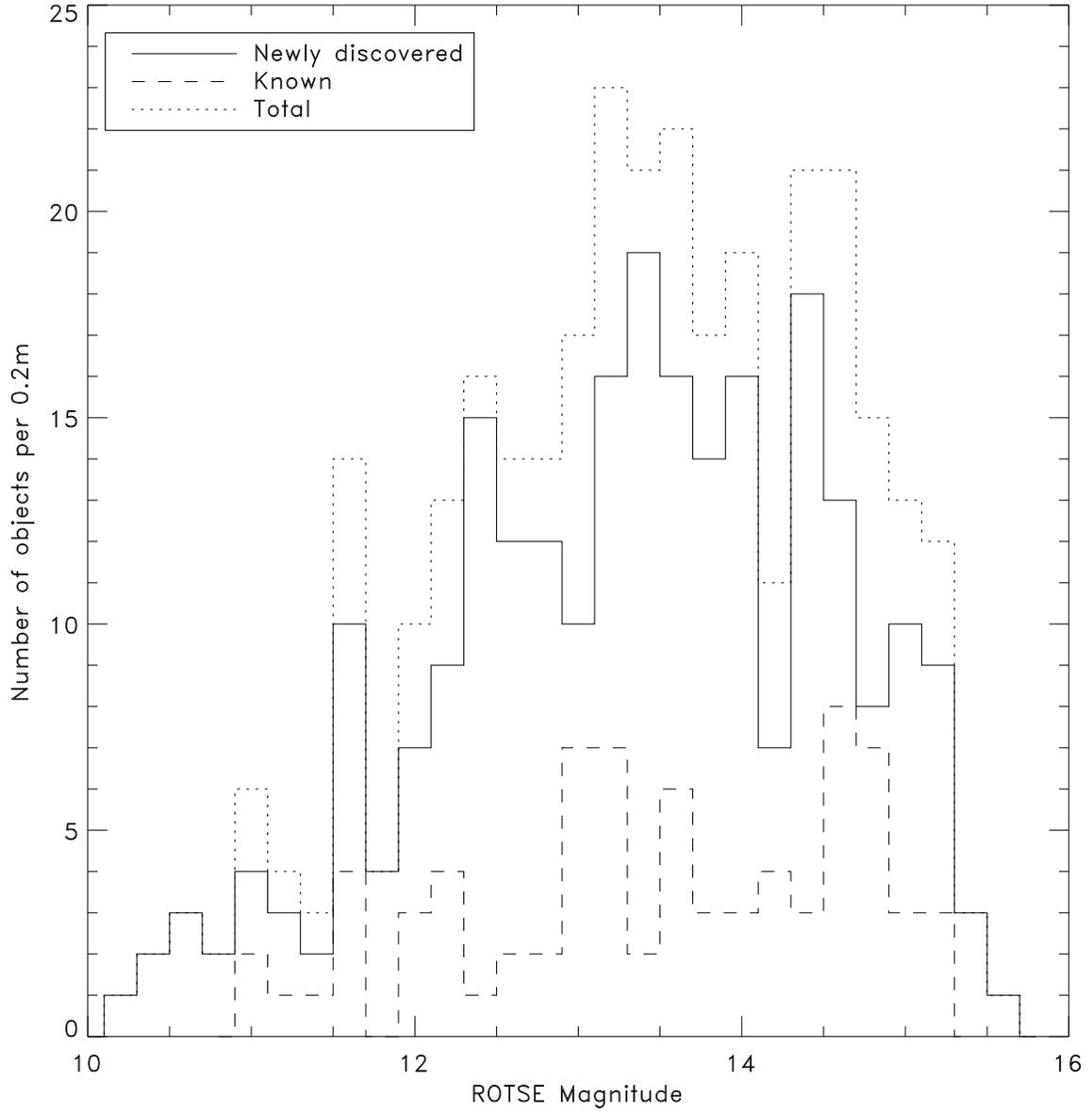}
\figcaption[rrl_mag_dist.ps]
{Magnitude distributions for RR Lyrae stars (RRAB and RRC) 
which are newly discovered 
(solid line), and previously known (dashed line). The dotted line
shows the magnitude distribution for the total sample.
\label{rrab_mag_dist}}
\end{figure}

RRC stars make up about 9\% of GCVS RR Lyraes. It is 
interesting to note that the RRC fraction found here, 38\%, is 
substantially larger. This illustrates the important advantage of 
ROTSE-I CCD photometry over earlier wide area surveys
based on photographic photometry. This difference is particularly striking
for variable classes like RRCs, with mean amplitudes $A_V$=0.3 m. 
The magnitude distributions for all new
and previously known RR Lyrae stars of both types  
are shown in Figure \ref{rrab_mag_dist}.

\subsection{Delta Scuti Stars}

The Delta Scuti stars are observationally (and physically)  
similar to the RRc class. They obey a period luminosity relation which 
is now well determined by Hipparcos calibration \citep{mcn97}.
Their periods range from 0.1 d to 0.28 d and their amplitudes range from
0.1 to 0.5 m. We have classified 91 objects as DS stars, of which
two are known from the GCVS. As with the RRC stars, the precision of
ROTSE CCD photometry helps to reveal large numbers of previously undetected
stars in this class. Examples of DS light curves are included
in Figure \ref{rrab_rrc_ds_lcplot}.

\subsection{Close Binary Systems}

\begin{figure}
\plotone{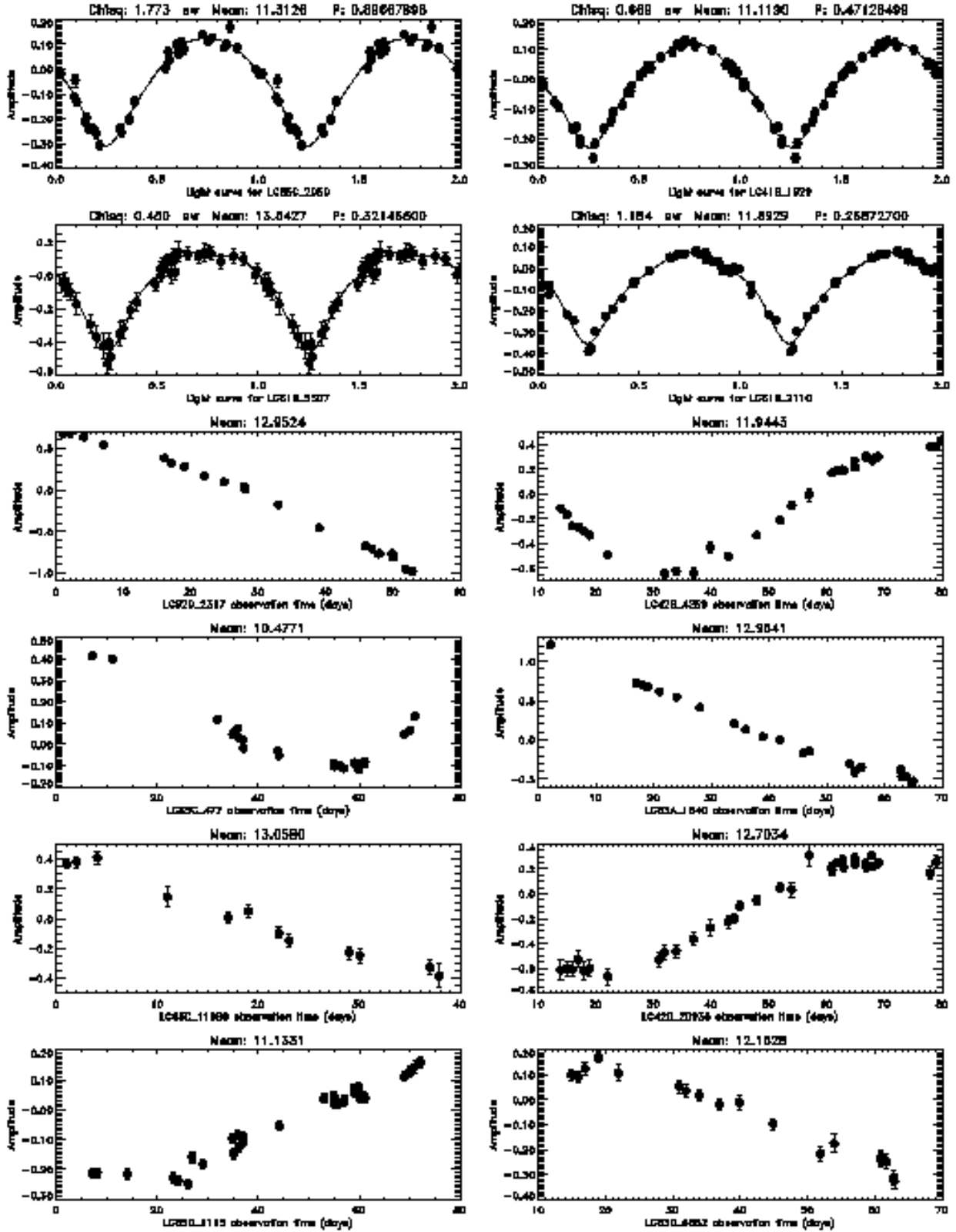}
\figcaption[ew_m_lpv_lcplot.ps]
{Example phased light curves for a random sample of ROTSE EW stars
(top four), along with unphased light curves for ROTSE M stars
(middle four) and ROTSE LPVs (bottom four). 
Included above each plot are periods (when phased)
and mean magnitudes. \label{ew_m_lpv_lcplot}}
\end{figure}

Close binary systems (mostly of the W UMa type) are very common in the 
ROTSE-I data. These objects have recently been shown to obey a reasonably
tight period-color-luminosity relation \citep{ruc97b}.
We have identified 382 candidate close binaries, of which
368 are new. The detection of such a large number of systems is not
unexpected. Most contact systems contain relatively low-mass G and K type
stars. It is estimated that as many as 1 in 500 G and K type stars are
members of contact binary systems.
Relatively shallow, wide area surveys such as this are
especially sensitive to such low luminosity
objects. 
Examples of EW systems are presented in Figure \ref{ew_m_lpv_lcplot}.

\subsection{Other Eclipsing Systems}

With between 22 and 114 epochs per location this analysis has relatively
poor sensitivity to widely separated eclipsing systems. Nonetheless, we
identify a total of 109 eclipsing systems, 95 of which are new. This is
a relatively inhomogeneous set, which includes both $\beta$ Lyrae systems
and detached Algol type eclipsing binaries.

\subsection{Intermediate period pulsators}

We have identified 201 systems with periods from 1 d to 100 d with more or
less sinusoidal light curves. All these objects are placed in class
C. They are not yet fully identified,
though it is clear that some are Cepheids, W Virginis stars, 
and RS CVn systems. 
Only 2 of these 201 objects are known in the GCVS; the dwarf nova 
AH Her, and HZ Her, a low-mass x-ray binary. An important application of 
the ROTSE all-sky variable survey will be identification of a complete
sample of bright Cepheids for calibration of the distance ladder.

\subsection{Miras and Other Long Period Variables}

The long period objects in our sample are drawn from at least two
different groups, Miras and red supergiant variables (RSVs). 
We have classified those with observed variations larger 
than one magnitude as M. 
There are 146 such objects in our sample, 66 of which are in the 
GCVS. Of the overlap objects, most (60 out of 66) are classified by GCVS as
Miras. The remainder are classified as long period or semi-regular 
variables. While periods for these very long period variables cannot be
firmly established by these data, we note that analysis of
the full two year ROTSE-I database will be extremely effective in this regard.

Miras are among the most venerable distance indicators, and obey a good 
period-luminosity relation, at least in K band \citep{bed98}. For
Miras this PL relation is contaminated in the optical by strong
TiO absorption. \citet{pie99} have recently shown that there is
a good optical PL relation for the red supergiant variables, and it
is likely that many of the objects classified here as M are actually 
in this class. Separation of these two classes should be possible using
IR data of the type which the 2MASS survey \citep{bei98} will provide.
As both types of stars are intrinsically very luminous 
($-4 <  M_{ROTSE} < -7$) they can be observed in ROTSE-I data to great
distance. The most luminous of these objects can be observed to 
about 300 kpc. While identification and period measurement of these
objects can be accomplished with existing ROTSE-I data, their use
as distance probes may require data in standard passbands.

We group all other long period variables into a single catch-all class. 
There are 534 such objects detected, of which 501 are newly discovered.
This large number of new objects is remarkable, especially in light of
our relatively low detection efficiency ($\sim$37\%) for these objects.
The combination suggests that longer duration observations will unveil
substantially larger numbers of LPVs.
Again, it is impossible within these four months of data to accurately
determine periods for these objects. Examples of M and LPV stars are 
presented in Figure \ref{ew_m_lpv_lcplot}. 

\section{Multiwavelength Correlations}

We have correlated our variable catalog with several all-sky catalogs
in other wavebands. Comparison to the ROSAT All Sky Survey Bright Source
Catalog \citep{vog99} yields 26 matches within a radius of 40''. Of these,
only four are listed in 
SIMBAD\footnote{\url{http://hea-www.harvard.edu/SIMBAD}};
AM Her (a CV), HZ Her (an LMXRB), 
PW HER (an RS CVn star) and TW CrB. The last is unidentified in SIMBAD,
but is clearly an EW object in ROTSE data. Of the remainder, 5 are
short period eclipsing systems, one exhibits a long ($>$100 d) fade
and 16 are longer period C type variables. These last may be RS CVn stars,
CVs, or x-ray binaries. We are planning a program of followup spectroscopy 
to determine the nature of these interesting objects.

We have also cross-correlated our catalog of variables with the IRAS Point
Source Catalog \citep{iras88}. A total of 269 matches are found within a 
radius of 40'' (which is a typical IRAS error ellipse major axis). Only 85
of these are known in the GCVS.
Every one of these objects is a long period 
variable, which we have classified as M (105), LPV (156), or C (8). 
All of these objects are detected in the IRAS 12 $\mu$m channel, and about
half are detected in the 25 $\mu$m channel. These characteristics are perfectly
consistent with their classification as pulsating giant stars.

\section{Conclusions and Future Prospects}

We have searched a small fraction of ROTSE-I sky patrol data for periodic 
variables using a robust selection technique. All detected variables 
have been phased and automatically classified, and the 
derived classifications have each been checked by visual 
inspection of the phased light curves. A large number of 
new variable stars are uncovered, including representatives of 
many known classes of periodic variables.

The most important conclusion of this work is confirmation that many 
new variables of all types await discovery at bright magnitudes, and
that the ROTSE-I data are a highly effective discovery archive. These data
allow accurate classification, period, and ephemeris determination for
all variables with amplitudes greater than $0.1\mbox{ m}$. We identify
a fraction 1781/917226 $\simeq$ 0.2\% of all observed objects as variable.
Because it is known that the variable selection used here is biased 
against some variable types this is a firm lower limit on the variability 
fraction.

ROTSE-I all sky patrol data are already having a real impact on the
study of bright variables. In the magnitude range from 10.5-12.5
the new variables presented here (973) represent a 30\% increase
in the total number of GCVS variables (3162) in this magnitude
range. This despite the fact that they are drawn from only 5\% of the
celestial sphere. 

We have confirmed the expectation that many variables of modest amplitude,
such as RRc and Delta Scuti stars, have escaped detection in earlier all
sky searches. We find that the fraction of RR Lyraes which are of the RRC
type is about 38\%, in contrast to the 9\% suggested by the GCVS.

The data analyzed here constitute only 5.6\% of the ROTSE-I sky coverage.
The entire sky visible from New Mexico has already been observed for 
a number of epochs ranging from 150 at -30\degr dec to 1100 at the 
north celestial pole. As this study 
includes substantial regions at both low (b $\sim$ 10\degr) and high 
(b=90\degr) galactic latitude, it is reasonable to predict the number
of objects we will detect in the full survey by simple extrapolation. From 
this extrapolation we expect to uncover a total of about 32,000 variable
stars via ROTSE-I patrol analysis.

Perhaps most important, ROTSE-I variable star selection will be carried 
out in a completely uniform way across the full sky north of 
-30\degr
dec. The entire catalog will be based on CCD photometry, obtained with 
a single instrument and reduced in a single, consistent way.
This will make the ROTSE-I variable catalog uniquely useful for 
studies of the galactic distribution of variables, and for direct studies 
of galactic structure.

We have concentrated here on the discovery and classification of periodic 
variables. Several selection criteria are deliberately biased against 
the inclusion of flaring objects. Analysis of these data for a variety
of aperiodic transients is underway, and will be reported in subsequent 
publications.

In addition to extending our analysis of existing data, the ROTSE program is
assembling a variety of new tools for the study of astrophysical transients.
In the near future we expect to begin repeating the ROTSE-I patrol 
scheme with modifications to allow us to obtain three color
(V, R, and I) observations. A fourth channel will retain
an open CCD for
increased sensitivity and comparison to earlier ROTSE-I results.
In the longer term, the ROTSE program is engaged in a large scale
expansion designed to improve our sensitivity to GRB optical counterparts.
We are constructing an array of ten 0.45m telescopes, each
with a 2\degr x 2\degr \hspace{0.4mm} field of view and a thinned 
CCD camera. These
telescopes will be deployed globally to a total of 6 sites, and will 
provide 24 hour coverage of the sky. These ROTSE-III instruments will allow
us to extend the kind of variable studies presented here to at least
19th magnitude while still covering a substantial fraction of the sky.







\acknowledgments

The authors acknowledge useful conversations with Bohdan Paczy\'{n}ski and 
Joyce Guzik, and thank Michael Pierce and John Jurcevic for making
available information on red supergiant variables prior to publication.
ROTSE is supported at the University of Michigan by NSF grants AST 9970818
and AST 9703282, NASA grant NAG5-5101, the Research Corporation, the 
University of Michigan, and the Planetary Society. Work performed at 
LANL is supported by the DOE under contract W7405-ENG-36, by NASA, and
by a Laboratory Directed Research and Development grant. Work performed
at LLNL is supported by the DOE under contract W7405-ENG-48.

\clearpage

\begin{deluxetable}{llllllllll}
\tablewidth{0pt}
\tabletypesize{\tiny}
\tablecaption{Example entries from the ROTSE-I catalog. 
\label{example_table}}
\tablehead{
\colhead{Name} & \colhead{Type} & \colhead{Ra} & \colhead{Dec} &
\colhead{\mR} & \colhead{$\sigma_{mR}$} & \colhead{$\Gamma$} &
\colhead{$\sigma_{\Gamma}$} & \colhead{Amp} & 
\colhead{Quality}}
\startdata
ROTSE1 J170711.48+454926.0 & rrab & 256.797852 & +45.823914 & 14.62 & 0.116 & 0.71538502 & 0.00042424 & 0.832 & 7 \\
ROTSE1 J174002.18+495318.9 & c  &  265.009125 & +49.888599 & 10.88 & 0.004 & 38.4667625 & 0.95186836 & 0.220 & 1 \\
ROTSE1 J172429.80+460801.7 & m  &  261.124207 & +46.133808 & 11.07 & 0.004 & 0.00000000 & 0.00000000 & 1.490 & 1 \\
ROTSE1 J171630.99+433832.1 & ew &  259.129150 & +43.642254 & 11.27 & 0.006 & 0.23350200 & 0.00003322 & 0.152 & 5 \\
\enddata
\tablecomments{These are sample entries from the catalog described by this
paper. The amplitudes reported here are derived from spline fits to the 
light curves. For objects with incomplete phase coverage these can be
inaccurate. Updated versions of the table, including new data, will be
occasionally made available at the ROTSE web site. [The complete version 
of this table is in the electronic edition of
the Journal.  The printed edition contains only a sample]}

\end{deluxetable}

\clearpage



\clearpage 

\begin{figure}
\figcaption[sky_patrol_map.ps]{Locations of ROTSE-I patrol
locations are shown with the galactic plane and familiar 
constellations overlaid. 
Regular patrol observations cover all available fields north of -30\degr
dec. \label{patrol_map}}
\end{figure}

\clearpage

\clearpage

\clearpage

\clearpage

\clearpage

\clearpage

\clearpage

\clearpage

\clearpage

\clearpage

\clearpage

\end{document}